\RequirePackage{xcolor}
\documentclass[journal,twoside, web]{ieeecolor}
\usepackage{tmi}
\usepackage[backend=biber,style=ieee,natbib=true,maxbibnames=9,maxcitenames=1,mincitenames=1,sorting=nty]{biblatex} 
\usepackage{makecell}
\usepackage{mathtools}

\usepackage{amsmath,amssymb,amsfonts,centernot}

\usepackage[noline,ruled,linesnumbered]{algorithm2e}

\SetCommentSty{mycommfont}

\newcommand{\mycomment}[1]{}

\usepackage{soul}
\usepackage{graphicx}
\usepackage{textcomp}
\def\BibTeX{{\rm B\kern-.05em{\sc i\kern-.025em b}\kern-.08em
    T\kern-.1667em\lower.7ex\hbox{E}\kern-.125emX}}

\usepackage{booktabs, multirow}

\usepackage{changepage,threeparttable} %

\usepackage{bookmark}

\usepackage{threeparttable}
\usepackage{adjustbox}
\newif\ifgeneratetestwarnings
\generatetestwarningstrue

\newif\ifsuppressclasswarnings
\suppressclasswarningstrue
\ifsuppressclasswarnings
    \usepackage{etoolbox}
    \usepackage{atbegshi}

    \pretocmd{\maketitle}{%
        \newcounter{titlepage}\setcounter{titlepage}{\value{page}}%
        \newcount\oldhbadness\oldhbadness=\hbadness\hbadness=10000
        \newdimen\oldhfuzz\oldhfuzz=\hfuzz\hfuzz=162pt%
        \newdimen\oldvfuzz\oldvfuzz=\vfuzz\vfuzz=22pt%
    }{}{Error.}

    \AtBeginShipout{%
        \ifnumequal{\value{page}}{\value{titlepage}}{%
            \global\hbadness=\oldhbadness%
            \global\hfuzz=\oldhfuzz%
            \global\vfuzz=\oldvfuzz%
        }{}%
    }

    \makeatletter
    \patchcmd{\@evenhead}{%
        \vbox{\color{subsectioncolor}\hrule height1pt width43pc depth0pt}%
    }{%
        \parbox[c][15pt][c]{\textwidth}{\color{subsectioncolor}\hrule height1pt width43pc depth0pt}%
    }{}{Error.}
    \patchcmd{\@oddhead}{%
        \vbox{\color{subsectioncolor}\hrule height1pt width43pc depth0pt}%
    }{%
        \parbox[c][15pt][c]{\textwidth}{\color{subsectioncolor}\hrule height1pt width43pc depth0pt}%
    }{}{Error.}
    \makeatother
\fi

\makeatletter
\let\NAT@parse\undefined
\def\@footnotecolor{red}

\define@key{Hyp}{footnotecolor}{%
 \HyColor@HyperrefColor{#1}\@footnotecolor%
}
\def\@footnotemark{%
    \leavevmode
    \ifhmode\edef\@x@sf{\the\spacefactor}\nobreak\fi
    \stepcounter{Hfootnote}%
    \global\let\Hy@saved@currentHref\@currentHref
    \hyper@makecurrent{Hfootnote}%
    \global\let\Hy@footnote@currentHref\@currentHref
    \global\let\@currentHref\Hy@saved@currentHref
    \hyper@linkstart{footnote}{\Hy@footnote@currentHref}%
    \@makefnmark
    \hyper@linkend
    \ifhmode\spacefactor\@x@sf\fi
    \relax
  }%
\makeatother

\usepackage{hyperref}
\hypersetup{
     colorlinks = true,
     linkcolor = red,
     citecolor= green,
     footnotecolor=blue,
}
\usepackage[utf8]{inputenc} %
\usepackage[T1]{fontenc}    %
\usepackage{hyperref}       %
\usepackage{url}            %
\usepackage{booktabs}       %
\usepackage{amsfonts}       %
\usepackage{nicefrac}       %
\usepackage{microtype}      %
\usepackage{xcolor}         %
\usepackage{lipsum}
\usepackage{graphicx}
\usepackage{wrapfig}
\usepackage{amsmath}

\title{MRExtrap: Longitudinal Aging of Brain MRIs using Linear Modeling in Latent Space}
\author{Jaivardhan Kapoor, Jakob H. Macke, and Christian F. Baumgartner
\thanks{This work was supported by the German Research Foundation (DFG) through Germany's Excellence Strategy (EXC-Number 2064/1, PN 390727645) and the German Federal Ministry of Education and Research (Tübingen AI Center, FKZ: 01IS18039).}
\thanks{JK and JHM are with Machine Learning in Science, University of Tübingen \& Tübingen AI Center, Germany (e-mail: jaivardhan.kapoor@uni-tuebingen.de, jakob.macke@uni-tuebingen.de). JK is a member of the International Max Planck Research School for Intelligent Systems (IMPRS-IS). JHM is also with the Department Empirical Inference, Max Planck Institute for Intelligent Systems, Tübingen, Germany. CFB is with the Cluster of Excellence: Machine Learning -- New Perspectives for Science, University of Tübingen, Germany, and Faculty of Health Sciences and Medicine, University of Lucerne, Switzerland (e-mail: christian.baumgartner@unilu.ch).}
\thanks{We thank Michael Deistler, Lisa Haxel, and Guy Moss for their feedback on the manuscript.}
}

\begin{document}
\maketitle

\thispagestyle{empty} %

\begin{abstract}
Simulating aging in 3D brain MRI scans can reveal disease progression patterns in neurological disorders such as Alzheimer's disease.
Current deep learning-based generative models typically approach this problem by predicting future scans from a single observed scan.
We investigate modeling brain aging via linear models in the latent space of convolutional autoencoders (MRExtrap).
Our approach, MRExtrap, is based on our observation that autoencoders trained on brain MRIs create latent spaces where aging trajectories appear approximately linear.
We train autoencoders on brain MRIs to create latent spaces, and investigate how these latent spaces allow predicting future MRIs through linear extrapolation based on age, using an estimated latent progression rate $\boldsymbol{\beta}$.
For single-scan prediction, we propose using population-averaged and subject-specific priors on linear progression rates.
We also demonstrate that predictions in the presence of additional scans can be flexibly updated using Bayesian posterior sampling, providing a mechanism for subject-specific refinement.
On the ADNI dataset, MRExtrap predicts aging patterns accurately and beats a GAN-based baseline for single-volume prediction of brain aging.
We also demonstrate and analyze multi-scan conditioning to incorporate subject-specific progression rates. 
Finally, we show that the latent progression rates in MRExtrap's linear framework correlate with disease and age-based aging patterns from previously studied structural atrophy rates. 
MRExtrap offers a simple and robust method for the age-based generation of 3D brain MRIs, particularly valuable in scenarios with multiple longitudinal observations. 

\end{abstract}


\section{Introduction}
\label{sec:intro}

\begin{figure}
    \centering
    \includegraphics[width=\columnwidth]{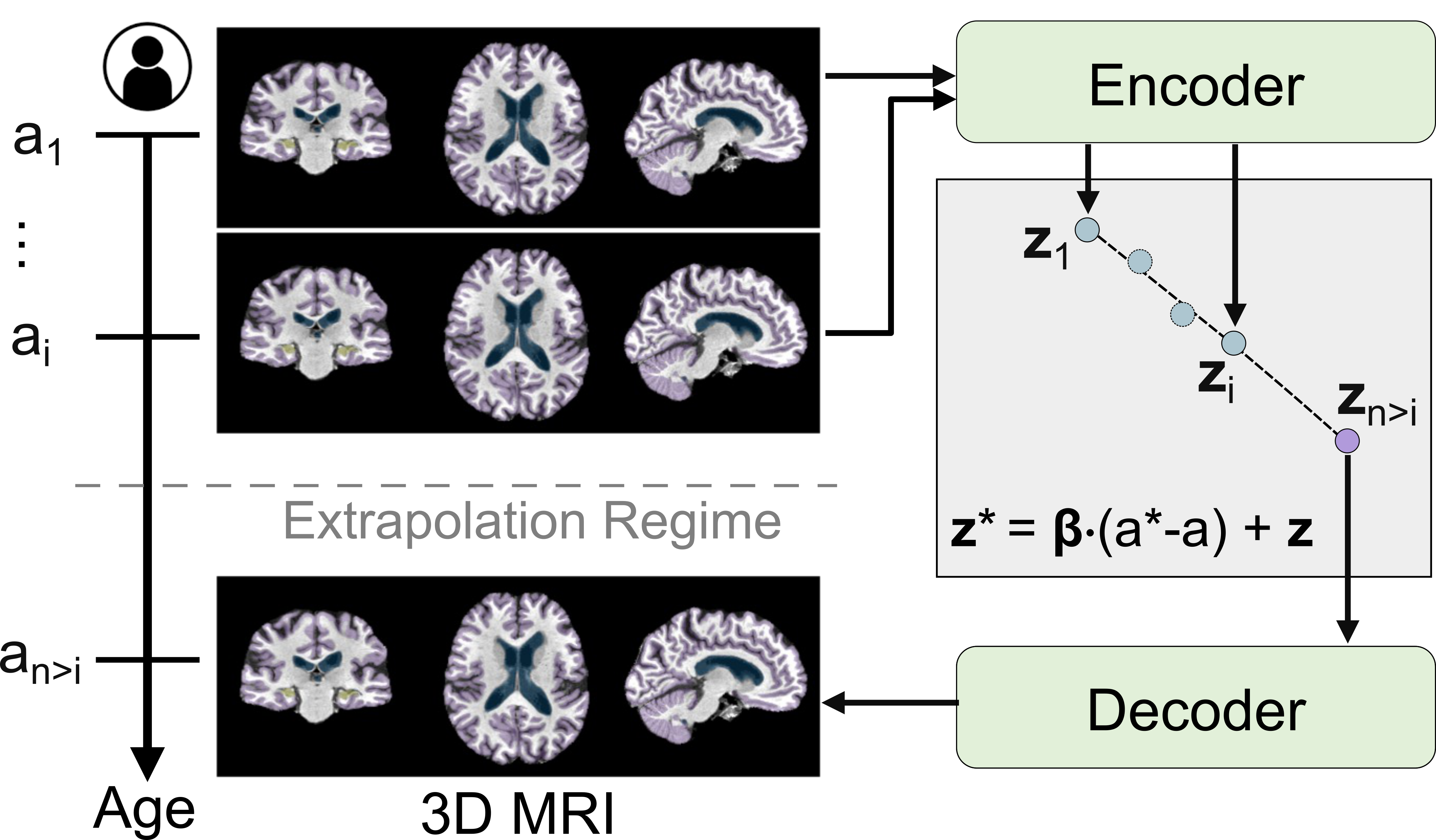}
    \caption{\textbf{MRExtrap models brain age progression as a linear model in the latent space of a 3D convolutional autoencoder.} We predict future brain MRIs from multiple observed scans by linearly extrapolating the latent representation $\mathbf{z}$ using a predicted/inferred progression rate $\boldsymbol{\beta}$. We overlay regional volumes (cortex, ventricles) in the MRI scans here with different colours.}
    \label{fig:teaser}
\end{figure}

Predicting age-related structural changes in brain MRIs is an important problem in neuroimaging research.
Accurate prediction of brain aging trajectories from neuroimaging data has significant implications for detecting anomalous brain atrophy patterns \cite{chouliaras2023use}, which may aid early diagnosis and monitoring of neurodegenerative diseases such as Alzheimer's disease.
Structural MRI scans, including T1, T1w, T2, and FLAIR modalities, offer non-invasive means of studying brain structure, and medical data consortia such as the Alzheimer's Disease Neuroimaging Initiative (ADNI) provide access to large databases of brain scans \cite{petersen2010alzheimer}.

Recent years have witnessed a growing body of work leveraging deep generative models for 3D brain generation \cite{vaegan,countersynth,ldm_kcl}, and in particular on structural aging \cite{daninet,vae_aging,sadm,jung2023conditional}.
These approaches typically employ deep learning architectures such as variational autoencoders (VAEs), generative adversarial networks (GANs), or diffusion models to capture the complex patterns of brain aging.
By learning from large datasets of longitudinal MRI scans, these models aim to generate plausible future brain states given one or more observed scans.
The potential applications of such models range from predicting individual disease trajectories to generating synthetic data for research and training purposes.

However, current approaches to generative modelling of 3D brain aging face several limitations:
Some methods face scalability issues to high resolutions \cite{sadm}, necessitating compromises such as slice-wise generation \cite{daninet}, that may not fully capture 3D structural changes.
Other methods involve training routines that require subject-specific finetuning~\cite{daninet, puglisi2024brlp}, making them challenging to implement and operate.
Existing models often operate in a pairwise setting, predicting from a single baseline scan, and lack the flexibility to incorporate a variable number of longitudinal observations at arbitrary ages of a subject.

Here, we investigate modeling brain aging via \textit{linear models in the latent space} of convolutional autoencoders.
Our approach, MRExtrap, builds on the observation that autoencoders trained on structural MRIs create latent spaces where the aging trajectories of subjects appear approximately linear.
Specifically, we first observe that within the latent space, regional brain volumes change linearly along interpolation paths between a subject's scans.
We then combine this with known linear aging patterns from neuroimaging literature to derive a direct linear relationship between age and latent representations (Sec.~\ref{sec:interp}).
Leveraging this linearity, MRExtrap predicts future MRI volumes via linear regression on sequences of these latent representations (by modeling the latent progression rate $\boldsymbol{\beta}$). We then decode the predicted latents back to voxel space after extrapolating for an arbitrary future age (Fig.~\ref{fig:teaser}).

Estimating the latent progression rate $\boldsymbol{\beta}$ is straightforward with multiple scans using linear regression, but is challenging when only a single baseline scan is available.
To address this, we propose priors on $\boldsymbol{\beta}$, ranging from a simple \textbf{Global Prior} derived from the average progression across the training set population, to learned, subject-specific amortized priors using UNet-parameterized Gaussian and Diffusion models (Sec.~\ref{sec:global_prior}, \ref{sec:amortized_priors}).
When additional longitudinal scans become available, MRExtrap can refine the $\boldsymbol{\beta}$ estimate using Bayesian posterior updating, allowing the model to incorporate subject-specific scan history (Sec.~\ref{sec:posterior_updating}).

We evaluate the MRExtrap framework on the ADNI dataset, and demonstrate that linear extrapolation using these priors accurately predicts structural aging patterns.
For single-scan prediction, the simple Global Prior proves remarkably effective and outperforms a GAN-based baseline \cite{daninet}.
We then analyze the effect of incorporating multiple scans via posterior updating to enable multi-volume conditioning, and find that its benefit depends on individual progression stability due to our linear model assumption.
Furthermore, we investigate the estimated latent progression rate $\boldsymbol{\beta}$ itself, finding that it meaningfully correlates with disease status and age, aligning with known patterns from volumetric atrophy studies.
Overall, this work presents a systematic investigation into linear latent space models for predicting voxel-level brain aging, offering MRExtrap as a simple, robust, and flexible framework applicable to longitudinal brain MRI data.

\section{Problem Setting}
\label{sec:problem}

To formally define the problem of predicting structural aging in brain MRIs (henceforth just referred to as brain aging), we consider a dataset $\mathcal{D} $ consisting of multiple subjects $s$, i.e., $\mathcal{D}=\{\mathcal{D}^{(s)}\}$, where $\mathcal{D}^{(s)}=\{(\mathbf{x}_i,a_i)\}_{i=1}^N$ is a sequence of $N$ 3D scans $\mathbf{x}_i\in\mathbb{R}^{D_d\times D_h\times D_w}$ recorded at ages $a_i\in\mathbb{R}^+$.
Here, the number of recorded scans $N$ can vary across subjects, and the ages $a_i$ are not necessarily uniformly spaced. For brevity, we omit the superscript $s$ denoting the subject.

For a given subject, our goal is to predict the scan $\mathbf{x}^*$ at an arbitrary target age $a^*>a_N$ given the observed scans $\mathcal{D}^{(s)}$ at ages $a_1\ldots a_N$, a conditional generative modeling task.
To this end, we first predict a low-dimensional latent representation $\mathbf{z}^*$ corresponding to age $a^*$, which is decoded back to obtain the predicted brain MRI $\mathbf{x}^*$.
This latent space is obtained via a convolutional autoencoder that compresses the 3D brain MRI $\mathbf{x}$ into a lower-dimensional volumetric latent representation $\mathbf{z}\in\mathbb{R}^{4\times D_d/f\times D_h/f\times D_w/f}$ (Section \ref{sec:methods}). 

\section{Autoencoder and linearity}
\label{sec:methods}
Our framework explores modeling brain aging within the latent space of a convolutional autoencoder. It involves two stages: compressing MRI volumes into this latent space, and then investigating linear models for age progression within it (see Fig.~\ref{fig:teaser}). In the following sections, we describe each stage, building up the MRExtrap approach.

\begin{figure*}[t]
    \centering
    \includegraphics[width=\textwidth]{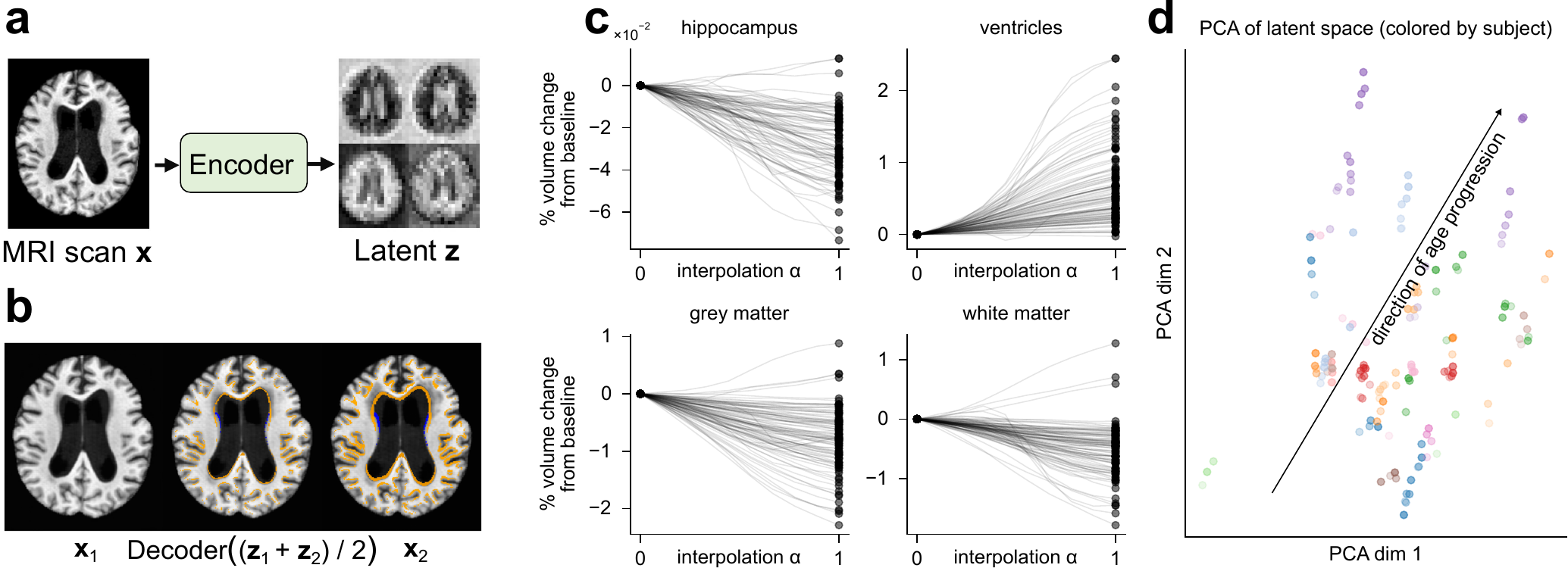}
    \caption{\textbf{Analysis of the latent space of the 3D convolutional autoencoder.} \textbf{a)} Latent codes show visual structures similar to those in the voxel space across all channels. \textbf{b)} Linearly interpolating the current and future latent representations within a subject, when decoded, results consistent volume changes with respect to current volume (overlayed in yellow). \textbf{(c)} Within each subject, volume changes across brain regions are roughly linear with respect to the interpolation factor $\alpha$. \textbf{(d)} PCA of latent codes, plotted across age (transparency) and subjects (colour). The age progression (indicated by the general direction of transparency) can be observed in the plot.}
    
    \label{fig:fig2interp}
\end{figure*}
\subsection{Perceptual compression using autoencoders}
\label{sec:autoencoder}

The compression of the MRIs into a latent space allows us to operate efficiently in a lower-dimensional space.
Following recent work on generative modelling \cite{van2017vqvae,ldm}, we train a convolutional autoencoder to compress 3D MRIs $\mathbf{x}$ to a latent representation $\mathbf{z}$. Similar to \citet{ldm} and \citet{ldm_kcl}, we use a combination of reconstruction losses, comprising an $L_1$ reconstruction loss, a loss based on structural similarity (SSIM), an LPIPS perceptual loss proposed by \citet{lpips}, and a patch-based adversarial discriminator loss AdvLoss \cite{patchgan}. Finally, we apply regularisation in latent space using the KL loss, obtaining the total loss 
\begin{align}
    \label{eq:loss}
    L_{\text{AE}} &= \Vert x-\hat{x} \Vert_1 + \text{SSIM}(x, \hat{x})+ \text{LPIPS}(x, \hat{x}) \\\nonumber
 & + \gamma_1\text{AdvLoss}(x,\hat{x}) + \gamma_2 D_{KL}(\text{Encoder}_{\mu, \Sigma}(x), \mathcal{N}(0,I)), 
\end{align}
where we set the loss scaling terms to $\gamma_1=0.005, \gamma_2=10^{-5}$, and $\hat{x}=\text{Dec}(\text{Encoder}_{\mu,\Sigma}(x))$.  Similar to a variational autoencoder, the latent representation $\mathbf{z}$ is modelled as a Normal distribution, and the $\text{Encoder}_{\mu,\Sigma}$ predicts its mean and standard deviation.
In practice, we only use the mean of the latent distribution for reconstruction after the network is trained, which we henceforth denote as $\mathbf{z}=\text{Encoder}(\mathbf{x})\in \mathbb{R}^{4\times D_d/f\times D_h/f\times D_w/f}$.

Since autoencoder-training only requires individual volumes, we can use cross-sectional as well as longitudinal data.
In line with previous work \citep{ldm, ldm_kcl}, the weight for the KL loss was set to a very small value to prevent the latent space from collapsing to an uninformative unit Gaussian distribution.
Our aim while training the autoencoder was to obtain highly accurate reconstructions of the input MRI volumes, which is crucial for subsequent modelling in the latent space. On the ADNI dataset, we obtained a mean SSIM of 0.95 and a mean Generalized Dice score \citep{sudre2017generaliseddice} of 0.92 (segmentation volumes of 4 brain regions using Synthseg+ \citep{synthsegplus}) between the real and reconstructed scans out of the autoencoder.

\subsection{Exploring Linearity in the Autoencoder Latent Space}
\label{sec:interp}

We analyzed the latent space of the trained autoencoder and found that it exhibits properties suggestive of local linearity, which motivates exploring linear models for aging.

\subsubsection{Linearity between latent codes and regional brain volumes}
\label{sec:linearity1}

The latent representations of the scans, due to their volumetric representation, preserve anatomical information relative to the voxel space (Fig.~\ref{fig:fig2interp}a).
We found that, within a subject, linearly interpolating between two latent representations $\mathbf{z}_1$ and $\mathbf{z}_2$ corresponding to corresponding to scans separated by approximately 2-4 years, produces decoded brain MRIs with regional volumes that appear to follow approximately linear interpolation.
This interpolation also preserves the subject-specific morphological structure for the hippocampus, ventricles, gray matter, and white matter volumes (Fig.~\ref{fig:fig2interp}b).
\begin{align*}
    \mathbf{z}_{\text{interp}} &= \alpha \cdot \mathbf{z}_1 + (1-\alpha) \cdot \mathbf{z}_2, \text{and}\\
    \hat{\mathbf{x}}_{\text{interp}} &= \text{Decoder}(\mathbf{z}_{\text{interp}}),
\end{align*}
the volumes of key brain regions obtained via segmentation of the interpolated brain images $\mathbf{v}_{\text{interp}} = \text{Seg}(\hat{\mathbf{x}}_{\text{interp}})$ generally follow an approximately linear relationship with the interpolation factor $\alpha$ (Figure~\ref{fig:fig2interp}c):
\begin{align}
    \mathbf{v}_\text{interp} \approx \alpha \cdot \mathbf{v}_1 + (1-\alpha)\cdot \mathbf{v}_2
\end{align}
We obtain the scalar volumes $\mathbf{v} \in \mathbb{R}^k$ (where $k$ is the number of brain regions) by segmenting the MRI scan using the SynthSeg$+$ software \cite{synthsegplus}, then compute the volume of the segmented regions.

This observed property implies that for each subject $s$, there might exist an approximate linear mapping between their latent representation $\mathbf{z}^{(s)}$ and their regional brain volumes $\mathbf{v}^{(s)}$:
\begin{align*}
    \mathbf{v}^{(s)} = C^{(s)} \cdot\mathbf{z}^{(s)} + \mathbf{c}^{(s)}
\end{align*}
where $C^{(s)}, \mathbf{c}^{(s)}$ are subject-specific variables. Omitting the subject superscript, the interpolation proceeds as follows:
\begin{align*}
    \mathbf{v}_{interp}&=\alpha\mathbf{v}_1+(1-\alpha)\mathbf{v}_2\nonumber\\
    &=\alpha(C \cdot\mathbf{z}_1 + \mathbf{c}) + (1-\alpha)(C \cdot\mathbf{z}_2 + \mathbf{c})\nonumber\\
    &=C \cdot (\alpha\mathbf{z}_1+(1-\alpha)\mathbf{z}_2) + \mathbf{c}=C \cdot\mathbf{z}_{interp} + \mathbf{c}
\end{align*}

While for small age differences voxel-space interpolation exhibits linearity in segmented volumes as well, it fundamentally treats MRIs as spatially aligned intensity maps.
This leads to blurred tissue boundaries and loss of structural detail under larger structural changes.
Latent-space interpolation instead preserves anatomical structures by operating on a low-dimensional manifold, and is computationally more efficient due to operating in a compressed representation space.

\subsubsection{Hypothesizing linearity between age and latent representations}

A broadly linear relationship between age and changes in scalar brain volumes has been observed in the literature using volumetric analyses \cite{salat2004thinning,fjell2009high, fujita2023characterization}, particularly for cortical thickness, ventricular volume, and hippocampus. This established relationship can be expressed as:
\begin{align*}
    \mathbf{v}(a) = B \cdot a + \mathbf{b}%
\end{align*}
where $B$ represents the rate of volume change with age, and $\mathbf{b}$ is the baseline volume.

We exploit these approximately linear relationships to model the progression of the latent space with age. Our reasoning is as follows: since regional brain volumes may be linearly modelled with respect to age, and latents within a subject appear to map linearly to brain volumes, we can hypothesize a direct linear relationship between age and latent representations.

From our two established relationships:
\begin{align*}
    \mathbf{v}(a) &= C \cdot\mathbf{z}(a) + \mathbf{c} \text{ \textcolor{gray}{(latent $\rightarrow$ volume)}}\\
    \mathbf{v}(a) &= B \cdot a + \mathbf{b} \text{ \textcolor{gray}{(age $\rightarrow$ volume)}}
\end{align*}

Combining these equations, we solve for $\mathbf{z}(a)$:
\begin{align*}
    C \cdot\mathbf{z}(a) &= B \cdot a + \mathbf{b} - \mathbf{c}\\
    \mathbf{z}(a) &= \underbrace{(C^+ \cdot B)}_{\boldsymbol{\beta}} \cdot a + \underbrace{C^+ \cdot (\mathbf{b} - \mathbf{c})}_{\mathbf{z}_0} \text{ \textcolor{gray}{(age $\rightarrow$ latent)}}
\end{align*}
where $C^+$ is the pseudoinverse of $C$. Here, $\boldsymbol{\beta}$ is the subject-specific \textit{progression rate} and $\mathbf{z}_0$ a subject-specific constant.

To examine this potential linear relationship, we projected the latents of all subjects onto a 2D space using PCA (Figure~\ref{fig:fig2interp}d). We observed that latents within a subject lie in an approximately linear subspace, and across subjects, the embeddings at later ages drift toward a consistent direction in the PCA space - the direction of age progression. This observation provides empirical support for exploring linear models.

\subsection{Linear Modeling of Brain Age Progression}

Based on the observations suggesting linearity between age $a\in\mathbb{R}^+$ and latents $\mathbf{z}\in\mathbb{R}^{4\times d\times d\times d}$, we can parametrize the linear model with a subject-specific progression rate $\boldsymbol{\beta}\in\mathbb{R}^{4\times d\times d\times d}$. For two latents $\mathbf{z}_i^{(s)}$ and $\mathbf{z}_j^{(s)}$ at ages $a_i, a_j$ for a subject $s$, this linear model can be written as:
\begin{align*}
    \mathbf{z}_i^{(s)}-\mathbf{z}_j^{(s)} &=  \boldsymbol{\beta}^{(s)}\cdot(a_i^{(s)}-a_j^{(s)}).
\end{align*}

If this linear model holds and we can estimate $\boldsymbol{\beta}^{(s)}$, we can predict $\mathbf{z}^*$ at a future age $a^*$. Given an estimate of the progression rate $\boldsymbol{\beta}^{* (s)}$ given two or more observed latents $\{ \mathbf{z}_i^{(s)}, a_i^{(s)}\}$, and setting the bias term using the most recent observation $\mathbf{z}^{(s)}_{N}$, the prediction is:

\begin{align*}
    \mathbf{z}^* &= \mathbf{z}^{(s)}_{N} + \boldsymbol{\beta}^{* (s)}\cdot(a^*-a_N^{(s)}).
\end{align*}

Fitting $\boldsymbol{\beta}^{* (s)}$ using standard linear regression requires at least two observed scans for a subject. This requires no fine-tuning of the model and is cheap to compute. By decoding the extrapolated latent $\mathbf{z}^*$ using the autoencoder's decoder, we can then obtain a subject-specific, voxel-level prediction of brain aging for an arbitrary future age.

\section{Modeling the progression rate $\boldsymbol{\beta}$}

The linear model described in Section \ref{sec:methods}-C relies on estimating the subject-specific progression rate $\boldsymbol{\beta}$.
While direct linear regression is possible with multiple scans, predicting from a single observation requires an alternative approach.
We propose probabilistic modeling of $\boldsymbol{\beta}$ using probabilistic methods, which allows brain aging using a single scan, while enabling us to incorporate future observations.
We consider a simple population-based Gaussian prior over the progression rate $\boldsymbol{\beta}$, as well as learned models for subject-specific priors. For the population-based Gaussian prior, we propose closed-form posterior updates for incorporating multiple observations.

\subsection{Computing the Progression Rate $\boldsymbol{\beta}$}
\label{sec:compute_beta}

To establish ground truth targets for learned models and to perform direct linear regression when possible, we first compute subject-specific progression rates from subjects with multiple observations in the training set. For each subject $s$ with MRI volumes at different ages:

\begin{align*}
    \mathbf{z}^{(s)}_i &= \text{Encoder}(\mathbf{x}^{(s)}_i) \quad \text{for each volume } i \\
    \Delta a &= [a^{(s)}_j - a^{(s)}_i \text{ for } j \neq i] \\
    \Delta \mathbf{z} &= [\mathbf{z}^{(s)}_j - \mathbf{z}^{(s)}_i \text{ for } j \neq i] \\
    \boldsymbol{\beta}^{(s)}_i &= \text{LinearRegression}(\Delta a, \Delta \mathbf{z}, \text{bias}=0)
\end{align*}

We use an L$_1$ loss function during regression to reduce sensitivity to outliers. This process generates a training dataset of triplets $(\mathbf{z}^{(s)}_i, a^{(s)}_i, \boldsymbol{\beta}^{(s)}_i)$.

\subsection{Global Linear Prior}
\label{sec:global_prior}

As a sensible starting point, we treat the progression rate for any subject as being drawn from a single, global distribution representing the average behavior of the training population. We term this the Global (linear) prior, which can be viewed as a Gaussian prior centered on the population mean progression rate.

We compute this prior from all the subject-specific progression rates $\boldsymbol{\beta}^{(s)}_i$ calculated in Section \ref{sec:compute_beta} across the entire training dataset $\mathcal{D}_\text{train}$.
The mean of the Global Prior is the element-wise average
\begin{align*}
    \boldsymbol{\mu}_\text{global} = \frac{1}{|\mathcal{D}_\text{train}|} \sum_{s,i \in \mathcal{D}_\text{train}} \boldsymbol{\beta}^{(s)}_i,
\end{align*}
and its variance is the diagonal empirical covariance $\boldsymbol{\Sigma}_\text{global}$, allowing the prior to be formally represented as $p(\boldsymbol{\beta}) = \mathcal{N}(\boldsymbol{\beta} | \boldsymbol{\mu}_\text{global}, \boldsymbol{\Sigma}_\text{global})$.
For single-scan prediction, we use the mean as a point estimate.

Prediction using this Global Prior involves applying the average progression rate $\boldsymbol{\mu}_\text{global}$ to the subject's latest scan $(\mathbf{z}_N, a_N)$:
\begin{align*}
    \mathbf{z}^* = \mathbf{z}_{N} + \boldsymbol{\mu}_\text{global} \cdot (a^* - a_N).
\end{align*}
This method is computationally trivial, requiring no subject-specific fitting beyond encoding the initial scan.
This model represents the simplest instantiation of the linear aging hypothesis averaged across the population.

\subsection{Amortized Priors for Single Observations}
\label{sec:amortized_priors}

Along with a single global prior for all subjects, we further explore learning probabilistic models that generate a subject-specific prior distribution $p_\theta(\boldsymbol{\beta}|\mathbf{z}, a)$, conditioned on the individual's baseline scan $(\mathbf{z}, a)$.
We explore two common classes of generative models for this task:

\subsubsection{UNet-based Gaussian Prior}
We model the conditional prior $p_\theta(\boldsymbol{\beta}|\mathbf{z}, a)$ as a Gaussian distribution.
A UNet architecture is trained to predict the parameters of this Gaussian -- the mean $\mu_\theta(\mathbf{z}, a)$ and the variance $\sigma^2_\theta(\mathbf{z}, a)$:
\begin{align*}
    p_\theta(\boldsymbol{\beta}|\mathbf{z}, a) = \mathcal{N}(\boldsymbol{\beta} | \mu_\theta(\mathbf{z}, a), \sigma^2_\theta(\mathbf{z}, a))
\end{align*}
The UNet is trained using the computed $\boldsymbol{\beta}^{(s)}_i$ from Section \ref{sec:compute_beta} as targets.
The outputs $\mu_\theta, \sigma^2_\theta$ from the UNet are trained to minimize the Gaussian negative loglikelihood $\mathcal{L}_\text{NLL} = \frac{\|\boldsymbol{\beta}^{(s)}_i - \mu_\theta\|^2}{\sigma^2_\theta} + \log \sigma^2_\theta$. We use an L$_1$ loss to handle outliers and use the combined loss $\mathcal{L}_{\text{UNet}}=\vert\mu_\theta- \boldsymbol{\beta}\vert_1 + 10^{-3}\mathcal{L}_\text{NLL}$ to train the UNet.

\subsubsection{Diffusion Model Prior}
We also train a denoising diffusion model \citep{ddpm} to implicitly learn the conditional prior distribution $p_\theta(\boldsymbol{\beta}|\mathbf{z}, a)$.
This allows potentially more complex and multi-modal distributions compared to the explicit Gaussian assumption.
The model trains a UNet-based denoiser network $\epsilon_\theta(\boldsymbol{\beta}_\text{noised} | \mathbf{z}, a, t)$ to predict the noise $\boldsymbol{\epsilon}$ added to a ground truth target $\boldsymbol{\beta}^{(s)}_i$ at diffusion timestep $t$.
We optimize the denoiser network's parameters by minimizing the mean squared error between the sampled and predicted noise, $\mathcal{L}_\text{diffusion} = \|\boldsymbol{\epsilon} - \hat{\boldsymbol{\epsilon}}\|^2$. 
To obtain a sample, we perform ancestral sampling starting from Gaussian noise $\boldsymbol{\beta}(T) \sim \mathcal{N}(0, I)$ using the standard reverse diffusion step:
\begin{align}
    \label{eq:diffusion_step}
    \resizebox{0.89\linewidth}{!}{$
    \boldsymbol{\beta}(t-1) = \frac{1}{\sqrt{\alpha_t}}\boldsymbol{\beta}(t)
    - \frac{1-\alpha_t}{\sqrt{\alpha_t(1-\bar{\alpha}_t)}}
    \,\epsilon_\theta\bigl(\boldsymbol{\beta}(t)\mid\mathbf{z},a,t\bigr) + \mathbf{n}
    $}
\end{align}
where $\alpha_t=\bar{\alpha}_{t}/\bar{\alpha}_{t-1}$ and $\mathbf{n}\sim \mathcal{N}(0, \frac{1-\bar{\alpha}_{t-1}}{1-\bar{\alpha}_t}(1-\alpha_t)I)$ is additional sampling noise.
At evaluation time, we generate $K=5$ samples of $\boldsymbol{\beta}$ from the diffusion model with different starting noises and average them to get a stable estimate.

These amortized models offer subject-specific priors from a single scan but involve computationally expensive training and inference compared to the Global Prior.

\subsection{Posterior Updating for Multiple Observations}
\label{sec:posterior_updating}

When additional longitudinal scans $\{(\mathbf{z}_j, a_j)\}$ become available for a subject, we can refine our probabilistic estimate of the progression rate $\boldsymbol{\beta}$. For a Gaussian prior over $\boldsymbol{\beta}$ $p(\boldsymbol{\beta}) = \mathcal{N}(\boldsymbol{\beta} | \mu_\text{pr}, \Sigma_\text{pr})$, the likelihood of observing a scan $\mathbf{z}_j$ at age $a_j$, given a rate $\boldsymbol{\beta}$ and the initial anchor point $(\mathbf{z}, a)$, is derived from the linear model: $\mathbf{z}_j \approx \mathbf{z} + \boldsymbol{\beta}(a_j - a)$.

Assuming Gaussian noise with variance $\sigma_\text{obs}^2$ for this relationship, the likelihood is $p(\mathbf{z}_j | \boldsymbol{\beta}, a_j, \mathbf{z}, a) \propto \exp(-\frac{1}{2\sigma_\text{obs}^2} \|\mathbf{z}_j - (\mathbf{z} + \boldsymbol{\beta}(a_j - a))\|^2)$.

For multiple independent observations $\{(\mathbf{z}_j, a_j)\}$, the combined likelihood multiplied by the Gaussian prior results in a Gaussian posterior distribution $p(\boldsymbol{\beta}|\mathbf{z}, a, \{(\mathbf{z}_j, a_j)\}) = \mathcal{N}(\boldsymbol{\beta} | \mu_\text{post}, \Sigma_\text{post})$, whose mean $\mu_\text{post}$ and covariance $\Sigma_\text{post}$ are given by the standard Bayesian linear regression update equations:
\begin{align}
\label{eq:unet_posterior}
    \mu_\text{post} &= \Sigma_\text{post} \left(\Sigma_\text{pr}^{-1}~ \mu_\text{pr} + \sum_j \frac{(a_j - a)  (\mathbf{z}_j - \mathbf{z})}{\sigma_\text{obs}^2}\right),\text{ where}\nonumber\\
    \Sigma_\text{post} &= \left(\Sigma_\text{pr}^{-1} + \sum_j \frac{(a_j - a)^2} {\sigma_\text{obs}^2}\right)^{-1}.
\end{align}

Finally, after obtaining an estimate $\boldsymbol{\beta}^*$ (which could be $\boldsymbol{\mu}_\text{global}$, $\mu_\theta$, a sample from the diffusion model, or $\mu_\text{post}$), we predict the latent representation at a future age $a^*$.
We use the \textit{most recent} observation $(\mathbf{z}_N, a_N)$ as the anchor point for extrapolation:
\begin{align*}
    \mathbf{z}^* &= \mathbf{z}_N + \boldsymbol{\beta}^* \cdot (a^* - a_N),\text{ then decoding to} \\
    \mathbf{x}^* &= \text{Decoder}(\mathbf{z}^*).
\end{align*}

\begin{table*}[t]
    \centering
    \scalebox{1.1}{
    \begin{tabular}{llcccc}
        \toprule
        \multicolumn{2}{c}{Method} & Hippocampus & Ventricle & Grey Matter & White Matter \\
        \midrule
        \multicolumn{2}{c}{DANINet \citep{daninet}}  & 0.060 $\pm$ 0.003 & \textbf{0.257 $\pm$ 0.017} & 0.829 $\pm$ 0.046 & 0.829 $\pm$ 0.046 \\
        \midrule
        \multirow{3}{*}{MRExtrap} & Global Gaussian Prior & \textbf{0.020 $\pm$ 0.001} & 0.322 $\pm$ 0.027 & \textbf{0.509 $\pm$ 0.030} & \textbf{0.399 $\pm$ 0.023} \\
        & UNet Gaussian Prior & \textbf{0.022 $\pm$ 0.001} & 0.310 $\pm$ 0.029 & 0.619 $\pm$ 0.043 & 0.443 $\pm$ 0.028 \\
        & Diffusion Prior & \textbf{0.021 $\pm$ 0.001} & \textbf{0.261 $\pm$ 0.019} & 0.639 $\pm$ 0.032 & \textbf{0.376 $\pm$ 0.022} \\
        \bottomrule
    \end{tabular}
    }
    \caption{
    Mean Absolute Error (MAE) with standard error across brain regions for 178 test subjects.
    MRExtrap is comparable to, or outperforms the baseline DANINet \cite{daninet} for predicting brain aging with a single observation.
    }
    \label{tab:single_vol_results}
\end{table*}

\section{Experimental and evaluation setup}
\label{sec:exp_setup}

\textbf{\color{mblue}{Data}~}
We used the Alzheimer's Disease Neuroimaging Initiative (ADNI) \citep{petersen2010alzheimer} database for our experiments.
We preprocessed 9,200 T1-weighted MRI scans from 1,700 patients, and split the dataset into 1411/32/178 subjects for training/validation/testing sets.
The average age of this cohort was 75 years, and we included subjects with all diagnoses. We considered the subject to be diagnosed as healthy if no element of the subject's sequence was diagnosed with mild cognitive impairment (MCI), nor with dementia. If the subject had a dementia diagnosis in the sequence, we considered the subject to be diagnosed with dementia. In all other cases, i.e., when the subject contained at least one MCI diagnosis, the subject was labeled as diagnosed with MCI. 
We kept the same testing set as used by \citet{daninet}.

The preprocessing steps included ROI clipping, bias field correction, affine registration (3+3 degrees of freedom) to the MNI152 atlas, and finally, skull stripping using \texttt{robex}.
This preprocessing ensures standardization across all scans and removes non-brain tissue, which is crucial for the accurate modeling of brain aging.
The preprocessed volumes have a resolution of $160\times 192\times 160$ mm$^3$ with voxel dimensions of 1mm$^3$.

\textbf{\color{mblue}{Training Details}~}
We trained an autoencoder with 6.6M parameters for 125 epochs using the AdamW optimizer \cite{loshchilov2017decoupled} with a learning rate of $10^{-4}$ and a cosine learning rate scheduler with a linear warmup.
To ensure training stability, we only applied the adversarial loss after 25 epochs, and linearly increased its weight, along with the KL loss weight, over one epoch.
The size of the latent space was set to $4\times 20\times24\times20$, resulting in a spatial reduction of $8\times$ in each dimension.

The UNet Gaussian Prior was parameterized with a 1.7M parameter UNet, trained to predict the mean $\mu_\theta$ and log-variance $\log\sigma^2_\theta$ of $\boldsymbol{\beta}$ given $(\mathbf{z}, a)$. We used the training triplets $(\mathbf{z}^{(s)}_i, a^{(s)}_i, \boldsymbol{\beta}^{(s)}_i)$ generated as described in Sec.~\ref{sec:compute_beta}. The UNet was trained for 2000 epochs using AdamW (learning rate $10^{-4}$, betas (0.95, 0.999)) with a cosine scheduler (4000 warmup steps) and the combined loss $\mathcal{L}_{\text{UNet}}$ defined in Sec.~\ref{sec:amortized_priors}.
We trained the diffusion model prior using an equivalent 1.7M parameter UNet denoiser $\epsilon_\theta$, and used the AdamW optimizer (learning rate $3\times10^{-4}$, betas (0.9, 0.999)) for 2000 epochs with a cosine scheduler (3000 warmup steps) on the MSE loss $\mathcal{L}_\text{diffusion}$. For inference, we used 500 diffusion steps with a linear noise schedule \citep{ddpm} and averaged over $K=5$ samples. We also applied an exponential moving average (decay 0.99) to model weights during training.
All models were trained and evaluated with NVIDIA A100 40GB GPUs.

For incorporating additional observations, we used the Global prior mean and variance using Eq.~\ref{eq:unet_posterior} with the observation noise variance $\sigma_\text{obs}^2$. We estimated this observation noise empirically from the training set by calculating the diagonal covariance of the residuals $\mathbf{z}_j - (\mathbf{z} + \boldsymbol{\beta}^{(s)}(a_j - a))$ over all subjects $s$ and timepoints $j$. Code is available at \url{http://github.com/mackelab/mrextrap}.

\textbf{\color{mblue}{Baselines}~}
To the best of our knowledge, MRExtrap is the first method that can consider more than two observed scans at arbitrary ages to predict brain aging.
Therefore, we compared our method in the single observation setting with DANINet, a GAN-based baseline proposed by \citet{daninet}.
We reported the single-volume predicted results directly from \citet{daninet}, which evaluated their method on $128\times 128$ cm$^2$ slices with 100 slices for each volume.  
For the multi-observation setting, we compare our amortization models to a linear regression baseline, which directly fits a linear model to the observed latents.

\textbf{\color{mblue}{Evaluation Metrics}~}
To assess the performance of our method, we used several evaluation metrics.
For volume prediction accuracy, we compared the predicted volumes of key brain regions (hippocampus, ventricular CSF, GM, and WM) with the actual volumes in held-out test scans. We report the volume changes as a percentage of the total brain volume (TBV) of the first seen scan.
For population-level comparisons, we computed the mean absolute error of the predicted vs.~real scans as a percentage of each subject's total brain volume at the first scan.
As in Sec.~\ref{sec:linearity1}, used the SynthSeg$+$ network \cite{synthsegplus} for the segmentation of both actual and predicted scans, and we reported the scalar volumes of these scans.

\begin{figure*}[t]
    \centering
    \includegraphics[width=\textwidth]{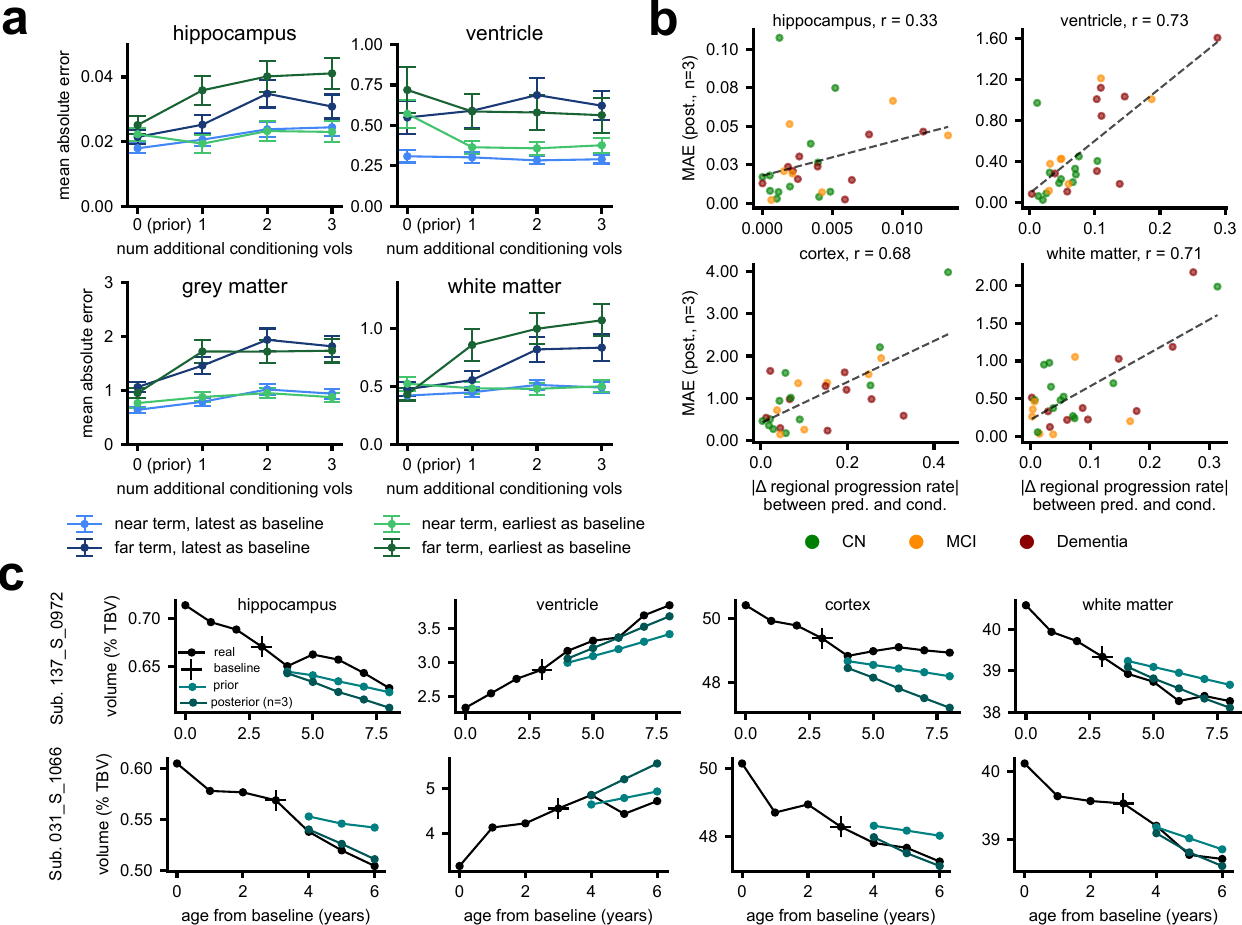}
    
    \caption{\textbf{Conditioning on multiple volumes reveals subject-specific and region-dependent updates to aging predictions.}
    \textbf{(a)} Mean absolute error (MAE, \% TBV) across 32 test subjects when conditioning on increasing numbers of past volumes (starting from the Global Prior at n=0). Error bars show standard error.
    \textbf{(b)} Subject-level MAE after conditioning on 3 volumes (posterior, n=3) plotted against the absolute change in regional progression rate ($\vert\Delta$rate$\vert$) between the conditioning (first 4y) and prediction (years 5+) periods.
    \textbf{(c)} Qualitative examples for two subjects comparing predictions using the Global Prior (light green) and the posterior after 3 observations (dark green).
    }
    \label{fig:multivol}
\end{figure*}

\section{Results}
\label{sec:results}

\begin{figure*}[t]
    \centering
    \includegraphics[width=\textwidth]{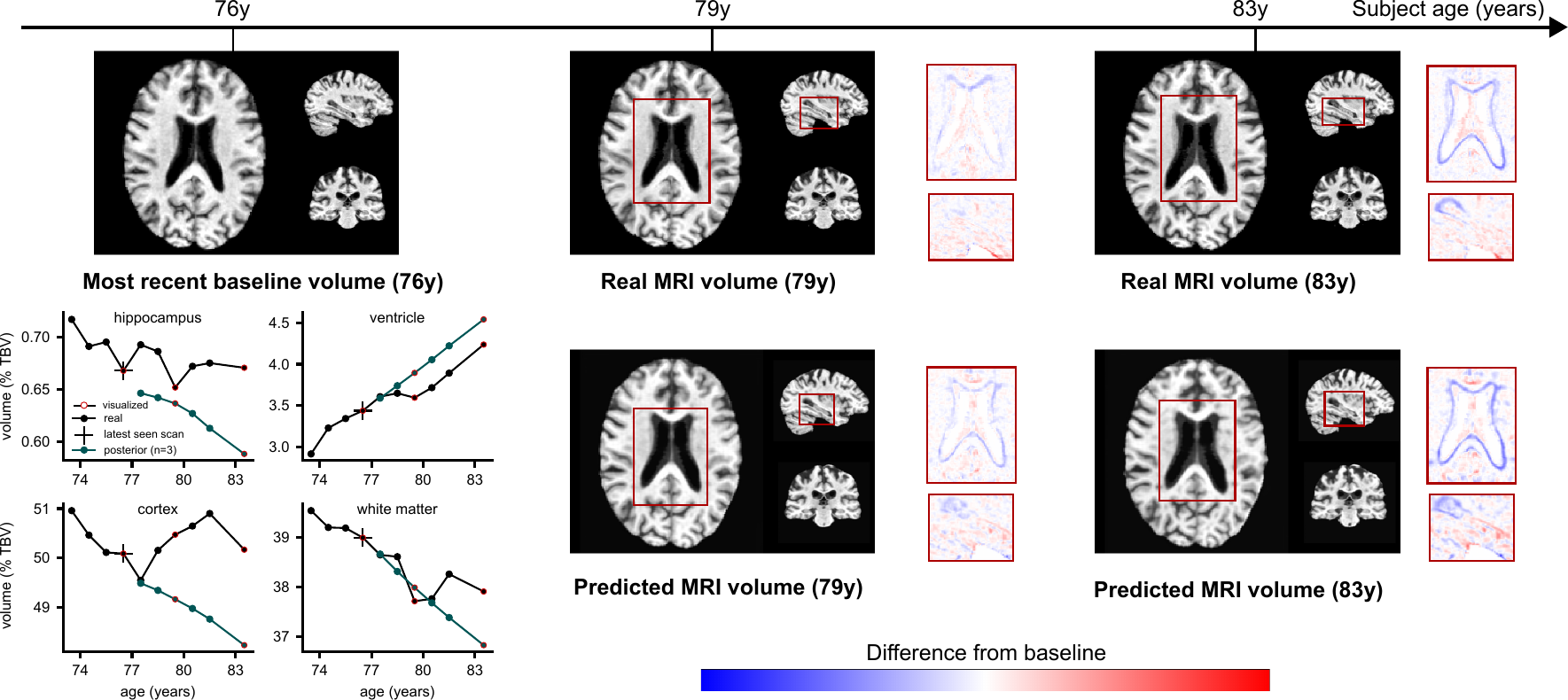}
    \caption{
    \textbf{MRExtrap predicts realistic structural aging patterns that maintain subject-specific anatomical integrity.} The top row shows the real brain MRI volumes at the target age for subject \texttt{137\_S\_0668}, while the bottom row shows the predicted volumes. The first four scans of the subject are used as input observations.
    }
    \label{fig:qualitative}
\end{figure*}

\subsection{MRExtrap accurately predicts brain aging patterns with a single observation}

\label{sec:results_single}
We first investigated the effectiveness of different linear progression priors for predicting future brain MRIs from a single observation.
For each subject in the test set (S=178), we chose a scan from the start of the longitudinal sequence.
We then predicted the brain at an age difference of 2 to 4 years and compared the predicted volumes with the actual volumes. We used the same initial and target age-volume pairs as in \citet{daninet} and compared MRExtrap with their reported numbers.

Remarkably, the simple Global Prior (Sec.~\ref{sec:global_prior}), which uses the average progression rate from the training set, achieves strong performance (Table \ref{tab:single_vol_results}).
It significantly outperforms the DANINet baseline in hippocampus, grey matter, and white matter regions ($p<0.01$, using reported DANINet mean/std err for z-test).
This result highlights that a basic linear model capturing the average population aging trend already provides a powerful baseline.

Comparing the Global Prior to the learned amortized priors (UNet Gaussian and Diffusion), we see relatively small differences---The UNet-parameterized Gaussian prior does not consistently outperform the Global Prior.
The Diffusion Prior shows comparable performance to the Global Prior for hippocampus and white matter, but yields improvements in ventricles ($p<0.1$ vs Global Prior, pairwise t-test). On the other hand, for grey matter, the Global Prior significantly outperforms the subject-specific priors.

Overall, these findings demonstrate that simple linear extrapolation in latent space, particularly using a straightforward Global Prior, is a surprisingly effective and simple method for single-scan brain aging prediction. As such, this Global Prior for linear extrapolation should be used as a strong baseline for future work in this field.
It remains unclear whether the added complexity of the learned amortized priors is worth the modest gains provided by subject-specific priors. We hypothesize that adapting the population-level estimate based on a single scan for a subject is inhertently difficult. Therefore, in subsequent multi-volume analysis, we use the Global Prior for updating the progression rate.

\subsection{Investigating Posterior Updates with Multiple Observations}
\label{sec:results_multi}

We next investigated whether incorporating additional longitudinal observations via posterior updating (Sec.~\ref{sec:posterior_updating}) improves prediction accuracy, implicitly testing the assumption of a constant linear progression rate over time.
We chose 24 subjects from the test set, which were scanned for at least 6 consecutive years. We considered the scan at 4 years from the first scan as the prior input, predicting the future scans in the sequence.
For additional observations, we considered scans lagging at 1 year, 2 years and 3 years, and the Global Prior $(\boldsymbol{\mu}_\text{global}, \boldsymbol{\Sigma}_\text{global})$ for the starting point for $\boldsymbol{\beta}$.
We then progressively incorporated scans from years 1, 2, and 3 using the Bayesian update rule (Eq.~\ref{eq:unet_posterior}) to obtain posterior estimates of $\boldsymbol{\beta}$.
Predictions were made for scans from year 4 onwards.

Fig.~\ref{fig:multivol}a shows the average MAE across these 24 subjects as more conditioning volumes are added.
Counterintuitively, on average, posterior updating does not consistently improve prediction accuracy compared to using the Global Prior alone.
While incorporating more scans reduces the standard error, the mean error often remains similar or even slightly increases compared to the 0-volume (Global Prior) baseline.

To understand this lack of average improvement on the population level, we examined the underlying regional progression rates, calculated from changes in regional scalar volumes over time.
Fig.~\ref{fig:multivol}b plots the correlation between the subject-level MAE, and the absolute change in regional progression rates (estimated via linear regression) between the conditioning period (first 4 years) and the subsequent prediction period (years 5 onwards) for each subject.
This analysis reveals variability in progression rates over time within individual subjects ($\vert\Delta$regional~progression~rate$\vert$), particularly evident in subjects with MCI/dementia (indicated by orange/red).
Our linear model assumes that $\boldsymbol{\beta}$ is constant, while the rate of change is often not constant across these multi-year intervals.
Consequently, a posterior estimate derived from an early time segment may not accurately reflect the progression rate during a later period.
On the other hand, the Global Prior, being an average over \textit{full} trajectories in the training set, may already capture the overall expected progression better than a posterior conditioned on a potentially non-representative initial trajectory segment.

This subject-specific variability leads to cases where the posterior update is beneficial, and others where it is detrimental, as shown by the qualitative examples in Fig.~\ref{fig:multivol}c. For subject \texttt{137\_S\_0972} (top), the posterior update (n=3, dark green) slightly improves the prediction for the ventricle and white matter compared to the Global Prior (light green), suggesting their progression rate was relatively stable during the observation period. On the other hand, due to abrupt changes in the regional progression rates in cortex (grey matter) and hippocampus, the posterior update diverges from the prior predictions. Conversely, for subject \texttt{031\_S\_1066}, the posterior enhances predictions for grey matter and hippocampus.

These findings suggest that while the linear model is a useful approximation, progression patterns in brain aging across long timescales ($\sim 10$y) may change rather than follow strictly constant trajectories.
Therefore, standard posterior updating, which relies on this assumption, does not consistently yield improved performance on average, although it can be beneficial on a case-by-case basis, assuming locally stable progression.

\subsection{Qualitative analysis of structural changes predicted by MRExtrap}
\label{sec:results_qualitative}

We considered three observed scans from the first 4 years of subject \texttt{137\_S\_0668}'s recording history from the test set, and predicted future observations using the posterior mean of $\boldsymbol{\beta}$ conditioned on these observations.
We observe that MRExtrap accurately predicts the structural changes in the brain, capturing the atrophy in the hippocampus and grey matter regions, and the expansion of the ventricles.
The volumetric plots show close alignment between real and predicted trajectories across all regions (Fig.~\ref{fig:qualitative}a).
The predicted MRIs demonstrate high fidelity to the real scans, with accurate prediction of ventricle expansion while maintaining overall brain structure (Fig.~\ref{fig:qualitative}b).

Importantly, MRExtrap preserves subject-specific anatomical structures in its predictions.
This preservation directly results from our linear latent space approach---since we're performing a linear extrapolation $\mathbf{z}^* = \mathbf{z} + \boldsymbol{\beta} \cdot (a^* - a)$, we maintain the core structural representation in $\mathbf{z}$ while adding only the age-related changes captured in $\boldsymbol{\beta}$.
The insets in Fig.~\ref{fig:qualitative}b highlight this preservation, showing that ventricle boundaries remain anatomically plausible and tissue interfaces maintain their realistic appearance.

\vspace{-0.5em}
\subsection{Behavior of of $\boldsymbol{\beta}$ across disease labels}

\begin{figure*}[t]
    \centering
    \includegraphics[width=1.0\textwidth]{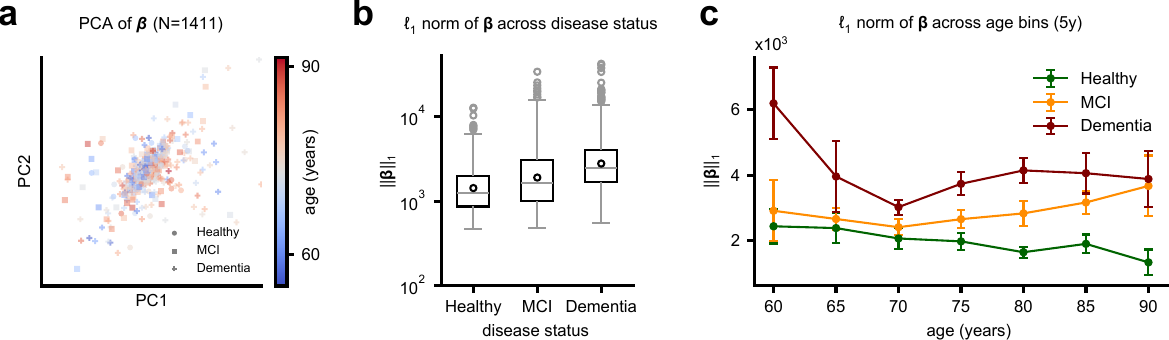}
    \caption{\textbf{Interpreting the progression rate $\boldsymbol{\beta}$ based on age and disease status.} 
    \textbf{(a)} The PCA of $\boldsymbol{\beta}$ inferred at the midpoint of each subject's full series in the training set shows outliers that have mild impairment or dementia disease labels.
    \textbf{(b)} $\Vert\boldsymbol{\beta}\Vert_1$, the $\ell_1$ norm of progression rate $\boldsymbol{\beta}$, differs in magnitude across disease labels.
    \textbf{(c)} When binned across age, $\Vert\boldsymbol{\beta}\Vert_1$ shows a clear separation between diseased and healthy subjects for all bins. Mild impairment lies between healthy subjects and subjects with dementia.
    }
    \label{fig:beta_mag}
\end{figure*}

We investigated whether the progression rate $\boldsymbol{\beta}$ contains age and disease-specific information across subjects (Fig.~\ref{fig:beta_mag}).
We computed $\boldsymbol{\beta}$ from the complete sequence for each subject in the training set at each observation. Overall, we found that $\boldsymbol{\beta}$, specifically its $\ell_1$ norm, behaves similarly to atrophy rates computed with volumetric analyses in the literature, and its behaviour across disease and age can be validated by previous studies \citep{leung2013cerebral, fiford2018patterns, schafer201predicting}.

We observed that plotting the first two principal components of $\boldsymbol{\beta}$ at the midpoint of each subject's sequence shows outliers that are either diagnosed with MCI or dementia (Fig.~\ref{fig:beta_mag}a).
In addition to PCA, we also plotted the distribution $\Vert\boldsymbol{\beta}\Vert_1$, the $\ell_1$ norm of $\boldsymbol{\beta}$, across diagnoses (Fig.~\ref{fig:beta_mag}b). We observed a clear increase in $\Vert\boldsymbol{\beta}\Vert_1$ from healthy to MCI, and from MCI to dementia group.
This observation aligns well with accelerated whole-brain atrophy rates in healthy vs. MCI vs. Alzheimer's subjects (\citep{leung2013cerebral} Tab.~2,\citep{schafer201predicting}).

To account for age- vs disease-related changes in the population, we sorted the subjects by the age of their first scan into 5-year interval bins. We then plotted the mean $\Vert\boldsymbol{\beta}\Vert_1$ across these bins for the different diagnosis groups (Figure \ref{fig:beta_mag}c). We observed that there is a clear difference in the norm across disease groups when accounting for age. In particular, in the younger age group of 60-70 years, we see that the norm is significantly higher for the dementia population than in later ages, which is in agreement with \citet{fiford2018patterns} (Sec.~3.2.2 of the study). While other changes across age may be visible with this binning (such as decreased norm for the healthy population at advanced ages), we attribute this to a sampling bias in subject screening in the ADNI database.

\vspace{-0.5em}
\section{Related Work}
\label{sec:related}

\subsection{Generative modeling of 3D brain MRIs}
Previous work on 3D brain generation has employed a variety of deep generative models, including generative adversarial networks (GANs) \cite{vaegan,vagan,countersynth,daninet}, variational autoencoders (VAEs) \citep{icamreg,vqvae_kcl,vae_aging,kapoor2023multiscale}, and more recently, diffusion models \cite{ddpm,diffusion_beats_gans,diffusion_beats_gans_medical,ldm}.
These approaches have shown promising results in synthesizing realistic brain MRIs, frequently conditioned on scalar attributes such as age, volume, or sex.
Notably, Latent Diffusion Models (LDMs) \cite{ldm}
have been applied to brain generation using the UK Biobank dataset, demonstrating effective conditioning on scalar attributes \cite{ldm_kcl}.
Models that operate in latent space, such as VAEs and latent diffusion models, offer increased efficiency and performance.

\subsection{Generative models for brain age progression}

For the more constrained task of longitudinal brain MRI generation, several methods have been proposed, each with its own strengths and limitations.
\citet{daninet} introduced DANINet which uses 2D GANs with biologically inspired spatiotemporal constraints to generate 2D slices that are assembled into 3D volumes.
Another conditional GAN-based approach by \citet{countersynth}, performed counterfactual synthesis on a given 3D scan, simulating various conditions including aging and gender counterfactuals.
Both approaches consider a single baseline scan for the aging task.%

With the increasing popularity of diffusion models, many approaches based on this model class have been proposed for longitudinal brain MRI generation.
\citet{sadm} introduce a model composed of a spatiotemporal vision transformer \cite{dosovitskiy2020vit} and a diffusion model to autoregressively predict a sequence of 3D brain MRIs.
This method, while considering a full longitudinal sequence, fails to accommodate variable intervals or missing observations and is computationally intensive as it operates directly on the voxel space.
\citet{puglisi2024brlp} proposed a diffusion-based approach for brain age progression, which involves a multi-stage approach for training and finetuning an LDM.
The method trains an unconditional LDM, which is then finetuned with a ControlNet \cite{zhang2023controlnet} using paired MRI data from a new subject.
This approach, while effective, requires finetuning for each new subject and lacks the interpretability provided by our linear modeling approach.

Although our work focuses on brain age progression of 3D structural MRIs, several related works operate using scalar biomarkers or shape features of key brain regions, particularly in the context of Alzheimer's disease.
\citet{koval2021ad} proposed a method for charting disease progression using parametric models.
Parametric approaches have also been developed that operate on shape data of regional volumes such as hippocampus \cite{bone17progression}.
While these methods are more interpretable owing to their parametric form, they cannot work on full 3D structural MRI data, limiting their ability to capture detailed spatiotemporal changes in brain structure.

\subsection{Linear interpolation in latent space}

MRExtrap makes heavy use of the linear relationships in the latent space of convolutional autoencoders.
The observation that linear interpolation in the latent space produces realistic and smooth variations among data points, is also studied in many generative model classes.
\citet{oring2020interp} and \citet{white2016sampling} study the effects of different interpolation methods in the latent space of autoencoders, and find that, generally, spherical interpolation performs better than linear interpolation due to the "soap bubble effect" \cite{huszar2017gaussian} in high-dimensional latent spaces.
Nevertheless, our analyses indicate that linear interpolation is predictive of structural MRI progression in several subjects---within individuals, the latents appear to lie approximately on a linear manifold, and a linear progression model is sufficiently expressive.

\vspace{-1em}

\section{Discussion}
\label{sec:discussion}

In this study, we explored MRExtrap for investigating the use of linear models for predicting brain aging patterns in 3D MRI scans.
MRExtrap leverages a convolutional autoencoder to compress MRIs into a latent space where aging trajectories appear approximately linear, enabling the exploration of (simple yet effective) linear extrapolation for predicting future scans.
We applied MRExtrap to model the brain age progression in subjects in the ADNI dataset, demonstrating its effectiveness and evaluating its limitations in predicting structural changes in key brain regions.

In single-scan predictions, we demonstrated that a simple global prior, derived from the average population progression rate, clearly outperforms more complex GAN-based baselines.
This highlights that linear latent modeling alone provides a robust and interpretable baseline.
Although learned, subject-specific priors showed modest improvements in some regions, the complexity added may not consistently justify the incremental gains.

MRExtrap assumes a subject-wise constant linear model of brain aging in the latent space, and the posterior updating mechanism specifically assumes constant linear progression rates within subjects.
However, our analysis also revealed that this latter assumption can break down, in particular among subjects with MCI or dementia, who exhibit substantial variability in their aging trajectories.
Specifically, we observed shifts in regional atrophy rates between the initial conditioning period and later prediction intervals (Fig.~\ref{fig:multivol}b,c).
Moreover, while the brain ages monotonically across decades, some nonlinearity may be observed in some regions. We observe the effect of this nonlinearity when conditioning our predictions on multiple scans.
Therefore, the limited performance gain from posterior updating is not necessarily a weakness of the method itself, but rather may result from inherent variability and non-stationarity in disease-driven aging patterns, as well as these potential underlying nonlinearities. 
Relaxing MRExtrap’s linear assumption by incorporating flexible, nonlinear models for latent progression could enhance expressiveness without sacrificing interpretability.

Despite these limitations, the progression rates themselves provide clinically meaningful signals, separating healthy from diseased populations.
This demonstrates their value as interpretable biomarkers in longitudinal aging studies.
MRExtrap is especially practical in scenarios demanding interpretability and efficiency, such as initial clinical assessments, low-resource settings, or when more complex nonlinear generative approaches prove unnecessary.
We see MRExtrap as a valuable baseline—transparent, interpretable, and computationally inexpensive—providing a reliable benchmark for more elaborate methods.

\printbibliography

@inproceedings{vaegan,
  author    = {Gihyun Kwon and Chihye Han and Daeshik Kim},
  title     = {Generation of 3D Brain {MRI} Using Auto-Encoding Generative Adversarial Networks},
  booktitle = {{MICCAI} 2019}
}

@inproceedings{vagan,
  author    = {Christian F. Baumgartner and Lisa M. Koch and Kerem Can Tezcan and Jia Xi Ang and Ender Konukoglu},
  title     = {Visual Feature Attribution Using Wasserstein GANs},
  booktitle = {2018 {IEEE} Conference on Computer Vision and Pattern Recognition, {CVPR} 2018, Salt Lake City, UT, USA, June 18-22, 2018},
  year      = {2018},
}

@article{countersynth,
  title   = {Equitable modelling of brain imaging by counterfactual augmentation with morphologically constrained 3D deep generative models},
  author  = {G. Pombo and R. Gray and M. Cardoso and S. Ourselin and G. Rees and J. Ashburner and P. Nachev},
  journal = {Medical Image Anal.},
  year    = {2021}
}

@article{daninet,
  title     = {Degenerative adversarial neuroimage nets for brain scan simulations: Application in ageing and dementia},
  author    = {Ravi, Daniele and Blumberg, Stefano B and Ingala, Silvia and Barkhof, Frederik and Alexander, Daniel C and Oxtoby, Neil and others},
  journal   = {Medical Image Analysis},
  volume    = {75},
  pages     = {102257},
  year      = {2022},
  publisher = {Elsevier}
}

@inproceedings{icamreg,
  title     = {{ICAM}-reg: Interpretable Classification and Regression with Feature Attribution for Mapping Neurological Phenotypes in Individual Scans},
  author    = {Cher Bass and Mariana da Silva and Carole H. Sudre and Logan Zane John Williams and Petru-Daniel Tudosiu and Fidel Alfaro-Almagro and Sean P. Fitzgibbon and Matthew Glasser and Stephen M. Smith and Emma Claire Robinson},
  booktitle = {Medical Imaging with Deep Learning},
  year      = {2021},
  
}

@inproceedings{vqvae_kcl,
  title        = {Morphology-Preserving Autoregressive 3D Generative Modelling of the Brain},
  author       = {Tudosiu, Petru-Daniel and Pinaya, Walter Hugo Lopez and Graham, Mark S and Borges, Pedro and Fernandez, Virginia and Yang, Dai and Appleyard, Jeremy and Novati, Guido and Mehra, Disha and Vella, Mike and others},
  booktitle    = {SASHIMI Worksop, MICCAI},
  year         = {2022}
}

@article{ldm_kcl,
  title     = {Brain Imaging Generation with Latent Diffusion Models},
  author    = {W. H. Pinaya and Petru-Daniel Tudosiu and J. Dafflon and P. F. D. Costa and Virginia Fernandez and P. Nachev and S. Ourselin and M. Cardoso},
  journal   = {DGM4MICCAI Workshop, MICCAI},
  year      = {2022},
}

@inproceedings{diffusion_beats_gans,
  author    = {Prafulla Dhariwal and Alexander Quinn Nichol},
  editor    = {Marc'Aurelio Ranzato and Alina Beygelzimer and Yann N. Dauphin and Percy Liang and Jennifer Wortman Vaughan},
  title     = {Diffusion Models Beat GANs on Image Synthesis},
  booktitle = {Advances in Neural Information Processing Systems},
  year      = {2021},
}

@article{diffusion_beats_gans_medical,
  title   = {Diffusion Probabilistic Models beat GANs on Medical Images},
  author  = {Gustav Müller-Franzes and Jan Moritz Niehues and Firas Khader and Soroosh Tayebi Arasteh and Christoph Haarburger and Christiane Kuhl and Tianci Wang and Tianyu Han and Sven Nebelung and Jakob Nikolas Kather and Daniel Truhn},
  year    = {2022},
  journal = {arXiv preprint arXiv: Arxiv-2212.07501}
}

@article{ldm,
  title   = {High-Resolution Image Synthesis with Latent Diffusion Models},
  author  = {Robin Rombach and Andreas Blattmann and Dominik Lorenz and Patrick Esser and Björn Ommer},
  year    = {2022},
  journal = {CVPR}
}

@inproceedings{sadm,
  title     = {SADM: Sequence-Aware Diffusion Model for Longitudinal Medical Image Generation},
  author    = {Yoon, Jee Seok and Zhang, Chenghao and Suk, Heung-Il and Guo, Jia and Li, Xiaoxiao},
  booktitle = {Information Processing in Medical Imaging},
  year      = {2023}
}

@inproceedings{vae_aging,
  title     = {Variational autoencoder for regression: Application to brain aging analysis},
  author    = {Zhao, Qingyu and Adeli, Ehsan and Honnorat, Nicolas and Leng, Tuo and Pohl, Kilian M},
  booktitle = {MICCAI 2019}
}

@inproceedings{ddpm,
  author    = {Jonathan Ho and Ajay Jain and Pieter Abbeel},
  editor    = {Hugo Larochelle and Marc'Aurelio Ranzato and Raia Hadsell and Maria{-}Florina Balcan and Hsuan{-}Tien Lin},
  title     = {Denoising Diffusion Probabilistic Models},
  booktitle = {NeurIPS},
  year      = {2020}
}

@article{lpips,
  title   = {The Unreasonable Effectiveness of Deep Features as a Perceptual Metric},
  author  = {Richard Zhang and Phillip Isola and Alexei A. Efros and Eli Shechtman and Oliver Wang},
  year    = {2018},
  journal = {arXiv preprint arXiv: 1801.03924}
}

@article{patchgan,
  title   = {Image-to-Image Translation with Conditional Adversarial Networks},
  author  = {Phillip Isola and Jun-Yan Zhu and Tinghui Zhou and Alexei A. Efros},
  year    = {2016},
  journal = {arXiv preprint arXiv: 1611.07004}
}

@article{chouliaras2023use,
  title     = {The use of neuroimaging techniques in the early and differential diagnosis of dementia},
  author    = {Chouliaras, Leonidas and O’Brien, John T},
  journal   = {Molecular Psychiatry},
  volume    = {28},
  number    = {10},
  pages     = {4084--4097},
  year      = {2023},
  publisher = {Nature Publishing Group UK London}
}

@article{petersen2010alzheimer,
  title     = {Alzheimer's disease Neuroimaging Initiative (ADNI) clinical characterization},
  author    = {Petersen, Ronald Carl and Aisen, Paul S and Beckett, Laurel A and others},
  journal   = {Neurology},
  volume    = {74},
  number    = {3},
  pages     = {201--209},
  year      = {2010},
  publisher = {AAN Enterprises}
}

@article{jung2023conditional,
  title     = {Conditional GAN with 3D discriminator for MRI generation of Alzheimer’s disease progression},
  author    = {Jung, Euijin and Luna, Miguel and Park, Sang Hyun},
  journal   = {Pattern Recognition},
  volume    = {133},
  pages     = {109061},
  year      = {2023},
  publisher = {Elsevier}
}

@article{salat2004thinning,
  author  = {Salat, David H. and Buckner, Randy L. and Snyder, Abraham Z. and Greve, Douglas N. and Desikan, Rahul S.R. and Busa, Evelina and Morris, John C. and Dale, Anders M. and Fischl, Bruce},
  title   = {{Thinning of the Cerebral Cortex in Aging}},
  journal = {Cerebral Cortex},
  volume  = {14},
  number  = {7},
  pages   = {721-730},
  year    = {2004}
}

@article{synthsegplus,
  author  = {Benjamin Billot  and Colin Magdamo  and You Cheng  and Steven E. Arnold  and Sudeshna Das  and Juan Eugenio Iglesias },
  title   = {Robust machine learning segmentation for large-scale analysis of heterogeneous clinical brain MRI datasets},
  journal = {Proceedings of the National Academy of Sciences},
  year    = {2023}
}

@article{dosovitskiy2020vit,
  title     = {An Image is Worth 16x16 Words: Transformers for Image Recognition at Scale},
  author    = {A. Dosovitskiy and L. Beyer and Alexander Kolesnikov and Dirk Weissenborn and Xiaohua Zhai and Thomas Unterthiner and M. Dehghani and Matthias Minderer and G. Heigold and S. Gelly and Jakob Uszkoreit and N. Houlsby},
  journal   = {International Conference On Learning Representations},
  year      = {2020}
}

@article{puglisi2024brlp,
  title   = {Enhancing Spatiotemporal Disease Progression Models via Latent Diffusion and Prior Knowledge},
  author  = {Lemuel Puglisi and Daniel C. Alexander and Daniele Ravì},
  year    = {2024},
  journal = {MICCAI}
}

@article{zhang2023controlnet,
  title     = {Adding Conditional Control to Text-to-Image Diffusion Models},
  author    = {Lvmin Zhang and Anyi Rao and Maneesh Agrawala},
  journal   = {IEEE International Conference on Computer Vision},
  year      = {2023},
  doi       = {10.1109/ICCV51070.2023.00355},
  bibsource = {Semantic Scholar https://www.semanticscholar.org/paper/efbe97d20c4ffe356e8826c01dc550bacc405add}
}

@article{koval2021ad,
  title     = {AD Course Map charts Alzheimer’s disease progression},
  author    = {Koval, Igor and B{\^o}ne, Alexandre and Louis, Maxime and Lartigue, Thomas and Bottani, Simona and Marcoux, Arnaud and Samper-Gonzalez, Jorge and Burgos, Ninon and Charlier, Benjamin and Bertrand, Anne and others},
  journal   = {Scientific Reports},
  volume    = {11},
  number    = {1},
  pages     = {8020},
  year      = {2021},
  publisher = {Nature Publishing Group UK London}
}

@inproceedings{bone17progression,
  author    = {B{\^o}ne, Alexandre
               and Louis, Maxime
               and Routier, Alexandre
               and Samper, Jorge
               and Bacci, Michael
               and Charlier, Benjamin
               and Colliot, Olivier
               and Durrleman, Stanley},
  title     = {Prediction of the Progression of Subcortical Brain Structures in Alzheimer's Disease from Baseline},
  booktitle = {Graphs in Biomedical Image Analysis, Computational Anatomy and Imaging Genetics},
  year      = {2017}
}

@article{kapoor2023multiscale,
  title   = {Multiscale Metamorphic VAE for 3D Brain MRI Synthesis},
  author  = {Jaivardhan Kapoor and Jakob H. Macke and Christian F. Baumgartner},
  journal = {NeurIPS Workshop on Medical Imaging},
  year    = {2022}
}

@inproceedings{oring2020interp,
  title     = {Autoencoder Image Interpolation by Shaping the Latent Space},
  author    = {Oring, Alon and Yakhini, Zohar and Hel-Or, Yacov},
  booktitle = {Proceedings of the 38th International Conference on Machine Learning},
  pages     = {8281-8290},
  year      = {2021},
  editor    = {Meila, Marina and Zhang, Tong},
  volume    = {139},
  series    = {Proceedings of Machine Learning Research},
  month     = {18-24 Jul},
  publisher = {PMLR},
}

@article{white2016sampling,
  title   = {Sampling Generative Networks},
  author  = {Tom White},
  year    = {2016},
  journal = {arXiv preprint arXiv: 1609.04468}
}

@misc{huszar2017gaussian,
  title        = {Gaussian distributions are soap bubbles},
  author       = {Huszar, Ferenc},
  year         = {2017},
  howpublished = {\url{https://www.inference.vc/high-dimensional-gaussian-distributions-are-soap-bubble/}}
}

@article{van2017vqvae,
  title   = {Neural discrete representation learning},
  author  = {Van Den Oord, Aaron and Vinyals, Oriol and others},
  journal = {NeurIPS},
  year    = {2017}
}

@article{fjell2009high,
  title     = {High consistency of regional cortical thinning in aging across multiple samples},
  author    = {Fjell, Anders M and Westlye, Lars T and Amlien, Inge and Espeseth, Thomas and Reinvang, Ivar and Raz, Naftali and Agartz, Ingrid and Salat, David H and Greve, Doug N and Fischl, Bruce and Dale, Anders M and Walhovd, Kristine B},
  journal   = {Cereb Cortex},
  volume    = {19},
  number    = {9},
  pages     = {2001--2012},
  year      = {2009},
}

@article{fujita2023characterization,
  author  = {Fujita, Shohei and Mori, Susumu and Onda, Kengo and Hanaoka, Shouhei and Nomura, Yukihiro and Nakao, Takahiro and Yoshikawa, Takeharu and Takao, Hidemasa and Hayashi, Naoto and Abe, Osamu},
  title   = {{Characterization of Brain Volume Changes in Aging Individuals With Normal Cognition Using Serial Magnetic Resonance Imaging}},
  journal = {JAMA Network Open},
  year    = {2023}
}

@article{loshchilov2017decoupled,
  title   = {Decoupled Weight Decay Regularization},
  author  = {I. Loshchilov and F. Hutter},
  journal = {International Conference on Learning Representations},
  year    = {2017}
}

@article{leung2013cerebral,
  title={Cerebral atrophy in mild cognitive impairment and Alzheimer disease: rates and acceleration},
  author={Leung, Kelvin K and Bartlett, Jonathan W and Barnes, Josephine and Manning, Emily N and Ourselin, Sebastien and Fox, Nick C and Alzheimer's Disease Neuroimaging Initiative},
  journal={Neurology},
  volume={80},
  year={2013}
}

@article{schafer201predicting,
title = {Predicting brain atrophy from tau pathology: a summary of clinical findings and their translation into personalized models},
author = {Amelie Schäfer and Pavanjit Chaggar and Travis B. Thompson and Alain Goriely and Ellen Kuhl},
journal = {Brain Multiphysics},
volume = {2},
year = {2021},
}

@article{fiford2018patterns,
  title={Patterns of progressive atrophy vary with age in Alzheimer's disease patients},
  author={Fiford, Cassidy M and Ridgway, Gerard R and Cash, David M and Modat, Marc and Nicholas, Jennifer and Manning, Emily N and Malone, Ian B and Biessels, Geert Jan and Ourselin, Sebastien and Carmichael, Owen T and others},
  journal={Neurobiology of aging},
  volume={63},
  pages={22--32},
  year={2018},
  publisher={Elsevier}
}

@article{sudre2017generaliseddice,
  title     = {Generalised Dice overlap as a deep learning loss function for highly unbalanced segmentations},
  author    = {C. Sudre and Wenqi Li and Tom Kamiel Magda Vercauteren and S. Ourselin and M. Jorge Cardoso},
  journal   = {DLMIA/ML-CDS@MICCAI},
  year      = {2017},
  bibSource = {Semantic Scholar https://www.semanticscholar.org/paper/2d8edc4e38bf9907170238726ec902cb3739393b}
}
\end{document}